\pdfoutput=1

\documentclass[aps,prd,reprint,showpacs,showkeys,superscriptaddress]{revtex4-1}
\usepackage{amsmath}
\usepackage{bm}
\usepackage{graphicx}
\usepackage[pdftex,colorlinks,hyperindex,plainpages=false,bookmarksopen,bookmarksnumbered,pdfusetitle]{hyperref}
\usepackage{mathrsfs}

\let\de=\partial
\let\eps=\epsilon
\let\vp=\varphi
\newcommand\dd{\text{d}}
\newcommand\imag{\text{i}}
\newcommand\La{\mathscr{L}}
\newcommand\Ha{H}
\newcommand\Ma{\mathscr{M}}
\newcommand\Da{\mathscr{D}}
\newcommand\gr[1]{\text{#1}}
\newcommand\he[1]{#1^\dagger}
\newcommand\vek[1]{\bm{#1}}
\newcommand\bra[1]{\langle{#1}|}
\newcommand\ket[1]{|{#1}\rangle}
\begin{document}

\title{Scattering amplitudes of massive Nambu--Goldstone bosons}

\author{Tom\'{a}\v{s} Brauner}
\email{tomas.brauner@uis.no}
\affiliation{Department of Mathematics and Physics, University of Stavanger, 4036 Stavanger, Norway}
\author{Martin F.~Jakobsen}
\email{martin.f.jakobsen@ntnu.no}
\affiliation{Department of Physics, Norwegian University of Science and Technology, 7491 Trondheim, Norway}

\begin{abstract}
Massive Nambu--Goldstone (mNG) bosons are quasiparticles whose gap is determined exactly by symmetry. They appear whenever a symmetry is broken spontaneously in the ground state of a quantum many-body system, and at the same time explicitly by the system's chemical potential. In this paper, we revisit mNG bosons and show that apart from their gap, symmetry also protects their scattering amplitudes. Just like for ordinary gapless NG bosons, the scattering amplitudes of mNG bosons vanish in the long-wavelength limit. Unlike for gapless NG bosons, this statement holds for \emph{any} scattering process involving one or more external mNG states; there are no kinematic singularities associated with the radiation of a soft mNG boson from an on-shell initial or final state.
\begin{center}
\emph{M.~F.~Jakobsen would like to dedicate this article to his parents (Bj\o rnar \& Jannicke), grandparents (Frode \& Marit and Per-Gunnar \& Solbj\o rg), grandaunt (Turid) and sister (Katrine). Thank you for your unwavering support and encouragement during my studies.}
\end{center}
\end{abstract}

\maketitle

%%%%%%%%%%%%%%%%%%%%%%%%%%%%%%%%%%%%%%%%%%%%%%%%%%%

\section{Introduction}
\label{sec:intro}

Spontaneous symmetry breaking is one of the most important concepts in modern quantum physics. It is responsible for a vast range of phenomena, ranging from superfluidity and ferromagnetism to the generation of masses of elementary particles. As a rule, it is associated with the presence of gapless quasiparticles in the spectrum of the system: the Nambu--Goldstone (NG) bosons.

Exact global symmetries are, however, rare in nature. When the spontaneously broken symmetry is not exact but merely approximate, the associated soft mode acquires a gap. Such modes are referred to as pseudo-NG (pNG) bosons. As a rule, the gap of a pNG boson depends not just on symmetry alone, but also on the details of the dynamics of the system.

It turned out only recently that under certain circumstances, the gap of a pNG boson \emph{is} determined exactly by symmetry~\cite{Nicolis:2012vf}. Namely, breaking an otherwise exact symmetry by coupling a chemical potential to one of its generators leads to pNG-like modes with a gap fixed by the symmetry algebra and the chemical potential alone, independently of the details of the underlying dynamics. Such modes have been dubbed massive NG (mNG) bosons~\cite{Watanabe:2013uya}. The list of currently known mNG bosons covers a range of systems from condensed-matter to high-energy physics, and includes (anti)ferromagnetic magnons in an external magnetic field, the neutral pion in a pion superfluid in dense quark matter, certain excitations of relativistic Bose--Einstein condensates~\cite{Watanabe:2013uya}, and the Kohn mode~\cite{Kohn:1961zz} corresponding to center-of-mass oscillations of Bose--Einstein condensates in harmonic traps~\cite{Ohashi:2017vcy}.

The story is further complicated by the fact that not \emph{all} pNG bosons stemming from explicit breaking of a symmetry by a chemical potential are mNG bosons~\cite{Watanabe:2013uya,Nicolis:2013sga}. Examples of such states are somewhat exotic but not too difficult to construct, the simplest one perhaps appearing in a system where a global $\gr{SO(3)}$ symmetry is completely spontaneously broken. Unlike the true mNG bosons, the presence of such states in a given system is, however, not guaranteed. We will revisit this case in Appendix~\ref{app:O(3)model}.

The goal of this paper is to investigate further properties of mNG bosons beyond the sole fact that their gap is fixed by symmetry. It is well known that ordinary NG bosons interact weakly at low energies. More precisely, barring special circumstances leading to a kinematic singularity, the scattering amplitude for a process involving a NG boson vanishes in the limit where the momentum of this NG boson goes to zero. This fact is usually referred to as Adler's zero, and has recently been re-investigated intensively in the context of a constructive approach to scattering amplitudes; see Refs.~\cite{Cheung:2014dqa,Low:2014nga,*Kallosh:2016qvo,Cheung:2016drk} for some relevant publications on the subject~\footnote{In condensed-matter physics, scattering amplitudes are a somewhat less important observable than in high-energy physics. However, scattering of spin waves in (anti)ferromagnets has been analyzed using the effective Lagrangian formalism for instance in Ref.~\cite{Hofmann:1998pp,*Gongyo:2016dzp}.}.

Here we show that mNG bosons share this property despite their gap. In fact, the nonzero gap protects them against the mentioned kinematic singularities so that the scattering amplitude for \emph{any} process involving a mNG external state vanishes as its momentum goes to zero.

The plan of the paper is as follows. In Sec.~\ref{sec:current}, we review the basic facts about mNG bosons. We also discuss to some extent how current conservation, which is crucial for establishing the existence of Adler's zero, is modified in the presence of a chemical potential. In the next two sections, we then warm up by analyzing in detail two concrete examples of systems featuring a mNG boson. The system described in Sec.~\ref{sec:antiferromagnet} captures the behavior of antiferromagnetic magnons in an external magnetic field. Its key advantage is that its relativistic kinematics is unaffected by the chemical potential, which only modifies the perturbative interactions of magnons. The example studied in Sec.~\ref{sec:kaon}, known from certain scenarios for kaon condensation in dense quark matter~\cite{Miransky:2001tw,*Schafer:2001bq}, features fully nonrelativistic kinematics despite its relativistic origin. It thus brings to light most of the subtleties that we will have to deal with in Sec.~\ref{sec:general}, where a general argument for Adler's zero in scattering amplitudes of mNG bosons is presented. Finally, in Sec.~\ref{sec:conclusions} we summarize our findings and give some concluding remarks. We also discuss to some extent the limit in which the momenta of \emph{two} NG or mNG bosons, participating in a scattering process, are taken to zero simultaneously.

%%%%%%%%%%%%%%%%%%%%%%%%%%%%%%%%%%%%%%%%%%%%%%%%%%%

\section{Massive Nambu--Goldstone bosons and current conservation}
\label{sec:current}

Following Ref.~\cite{Watanabe:2013uya}, consider a quantum system defined by its Hamiltonian $\Ha$. Suppose that we pick one of the generators $Q$ of its symmetry group $G$ and assign it a chemical potential, $\mu$. The excitation spectrum of the system is then determined by the many-body Hamiltonian $\tilde\Ha\equiv\Ha-\mu Q$. This Hamiltonian generally does not commute with the full group $G$; let us denote the subgroup of $G$ commuting with $\tilde\Ha$ as $\tilde G$.

By the standard Cartan decomposition of Lie algebras, the symmetry generators not commuting with $\tilde\Ha$ can be organized into Hermitian-conjugate pairs $Q^\pm_i$ such that
\begin{equation}
[Q,Q^\pm_i]=\pm q_iQ^\pm_i,
\label{Qcommutator}
\end{equation}
where $q_i$ are the roots of the Cartan subalgebra.
It then follows that acting with $Q^\pm_i$ on an eigenstate of $\tilde\Ha$ changes its energy (eigenvalue of $\tilde\Ha$) by $\mp\mu q_i$. As a consequence, once both $\mu$ and $q_i$ are chosen without loss of generality to be positive, the many-body ground state $\ket0$ satisfies $Q^+_i\ket0=0$. On the other hand, $Q^-_i\ket0$ can be nonzero, and if it is (which signals spontaneous symmetry breaking), it represents a mNG state with energy $\mu q_i$.

The total number of mNG states in the spectrum can be determined as follows~\cite{Watanabe:2013uya}. Define the real antisymmetric matrix of commutators,
\begin{equation}
\rho_{ij}\equiv-\imag\lim_{\Omega\to\infty}\frac1\Omega\bra0[Q_i,Q_j]\ket0
\end{equation}
($\Omega$ denotes the spatial volume of the system), and the analogous matrix $\tilde\rho_{ij}$, composed of generators of $\tilde G$ only. The number of mNG bosons is then given by
\begin{equation}
n_\text{mNG}=\frac12(\text{rank}\,\rho-\text{rank}\,\tilde\rho).
\label{counting}
\end{equation}

To provide a somewhat different perspective on the spectrum of mNG bosons, we now discuss the conservation laws for Noether currents in the presence of a chemical potential. We use the fact that in the Lagrangian formalism, the chemical potential can be introduced as a constant background temporal gauge field~\cite{Kapusta:1981aa}.

Consider rather generally a class of theories defined by their classical action $S[\phi,A]$, depending on a set of scalar fields $\phi^a$ and gauge fields $A^i_\mu$. Suppose that this action is invariant under a set of simultaneous local transformations with infinitesimal parameters $\eps^i(x)$,
\begin{equation}
\delta\phi^a=\eps^iF_i^a(\phi,A),\qquad
\delta A^i_\mu=\de_\mu\eps^i+f^i_{jk}A^j_\mu\eps^k,
\label{gaugetransfo}
\end{equation}
where $f^i_{jk}$ are the structure constants of the symmetry group and $F_i^a$ some local functions of the fields and possibly of their derivatives. The requirement of gauge invariance implies the condition
\begin{equation}
\int\dd x\left[\frac{\delta S}{\delta\phi^a}\eps^iF_i^a+\frac{\delta S}{\delta A^i_\mu}(\de_\mu\eps^i+f^i_{jk}A^j_\mu\eps^k)\right]=0.
\end{equation}
By using the equation of motion for the \emph{scalar} field, $\delta S/\delta\phi^a=0$, we infer immediately that the Noether currents, defined by $J^\mu_i(x)\equiv\delta S/\delta A^i_\mu(x)$ up to a conventional sign, satisfy the covariant conservation law
\begin{equation}
D_\mu J^\mu_i\equiv\de_\mu J^\mu_i+f^k_{ij}A^j_\mu J^\mu_k=0.
\label{noether}
\end{equation}
Note the generality of our argument. First, we did not assume any particular form of the transformation rule for the scalar fields: the function $F^a(\phi,A)$ need not be linear, and it may even depend on field derivatives. Second, we did not make any specific assumptions on the Lagrangian density: it may depend on higher derivatives of the fields, and it may change upon the transformation~\eqref{gaugetransfo} by a surface term. Finally, the gauge field $A^i_\mu$ in Eq.~\eqref{noether} is treated as a non-dynamical background, but it may take an arbitrary coordinate-dependent value.

What we are actually interested in is the situation in which the background gauge field $A^Q_\mu$ for the generator $Q$ equals $(\mu,\vek0)$; all the other background gauge fields $A^i_\mu$ can be set to zero upon taking the functional derivative in order to obtain the Noether currents. It follows that the currents $J^\mu_\pm$ associated with the generators $Q^\pm$, satisfying Eq.~\eqref{Qcommutator} (we drop for the sake of simplicity the index $i$), fulfill the conservation law
\begin{equation}
\de_\mu J^\mu_\pm\pm\imag\mu qJ^0_\pm=0.
\label{conservation}
\end{equation}

Consider now the one-particle state of a mNG boson carrying momentum $\vek p$, denoted as $\ket{G(\vek p)}$. By the argument following Eq.~\eqref{Qcommutator}, this state can be created  from the many-body vacuum $\ket0$ by $Q^-$. The matrix element $\bra{G(\vek p)}J^\mu_-(x)\ket0$ is therefore nonzero. Spacetime translation invariance and spatial rotation invariance constrain it to take the form
\begin{equation}
\bra{G(\vek p)}J^\mu_-(x)\ket0=e^{\imag p\cdot x}\bigl[\imag p^\mu F_1(|\vek p|)+\imag\delta^{\mu0}F_2(|\vek p|)\bigr],
\label{FGdef}
\end{equation}
where $F_1(|\vek p|)$ and $F_2(|\vek p|)$ are a priori unknown functions of the mNG boson momentum. Applying the conservation law~\eqref{conservation} to the current $J^\mu_-$ then gives
\begin{equation}
\omega^2F_1+\omega(F_2-\mu qF_1)-\vek p^2F_1-\mu qF_2=0,
\end{equation}
where $\omega(\vek p)$ is the dispersion relation of the mNG mode. It is easy to see that $\omega(\vek 0)=\mu q$ is a solution of this equation for arbitrary $F_1$ and $F_2$, which provides yet another derivation of the mass of the mNG boson.

We shall utilize the conservation law~\eqref{conservation} and the matrix element~\eqref{FGdef} in our discussion of the mNG boson scattering amplitudes in the next sections.

%%%%%%%%%%%%%%%%%%%%%%%%%%%%%%%%%%%%%%%%%%%%%%%%%%%

\section{Case study: antiferromagnet in external magnetic field}
\label{sec:antiferromagnet}

Let us start our discussion of scattering amplitudes of mNG bosons by looking at a concrete example. It is clear from Eq.~\eqref{Qcommutator} that the presence of a mNG boson requires non-Abelian symmetry. We therefore choose to study the simplest non-Abelian relativistic model with the symmetry-breaking pattern $\gr{SO(3)}\to\gr{SO(2)}$. At the leading order of the derivative expansion, its low-energy effective Lagrangian is just the nonlinear sigma model,
\begin{equation}
\La=\frac12(D_\mu\vec\phi)^2,
\end{equation}
where the vector field $\vec\phi$ has a fixed modulus, $|\vec\phi|=v$. The covariant derivative includes a background gauge field $\vec A_\mu$ of $\gr{SO(3)}$ via
\begin{equation}
D_\mu\vec\phi\equiv\de_\mu\vec\phi+\vec A_\mu\times\vec\phi.
\end{equation}
For future reference, we take note of the Noether currents arising from the $\gr{SO(3)}$ symmetry,
\begin{equation}
\vec J_\mu=\frac{\delta S}{\delta\vec A_\mu}=\vec\phi\times D_\mu\vec\phi.
\label{noetherSO(3)}
\end{equation}
This model can be thought of as describing the low-energy dynamics of spin waves in antiferromagnets in an external magnetic field, represented by $\vec A_0$.

We choose the magnetic field to point along the $z$-axis, that is, set $\vec A_\mu=\delta_{\mu0}(0,0,\mu)$. In the classical ground state, the field $\vec\phi$ is then oriented in the $xy$ plane, and we can choose it to point in the $x$-direction, $\langle\vec\phi\rangle=(v,0,0)$. The fluctuations above this ground state are parameterized by two scalar fields, which we denote as $\pi$ and $G$ for a reason that will be clear shortly. We shall use the following nonlinear parameterization that automatically takes account of the constraint on the length of $\vec\phi$,
\begin{equation}
\vec\phi=\bigl(\sqrt{v^2-\pi^2-G^2},\pi,G\bigr)^T.
\end{equation}
Inserting this into the Lagrangian, it acquires a form that is suitable for a perturbative analysis of the model,
\begin{align}
\label{ferroLag}
\La={}&\frac12(\de_\mu\pi)^2+\frac12(\de_\mu G)^2-\frac12\mu^2G^2\\
\notag
&+2\mu(\de_0\pi)\sqrt{v^2-\pi^2-G^2}+\frac12\frac{(\pi\de_\mu\pi+G\de_\mu G)^2}{v^2-\pi^2-G^2},
\end{align}
up to constant and surface terms. We can see that the model contains one exactly massless mode and one mode with the mass equal to $\mu$, which is our mNG boson. This corresponds to the well-known fact that out of the two magnons in antiferromagnets, only one becomes gapped when an external magnetic field is turned on. Our notation then is: $\pi$ for the truly massless (NG) mode, and $G$ for the gapped (mNG) mode.

%%%%%%%%%%%%%%%%%%%%%%%%%%%%%%%%%%%%%%%%%%%%%%%%%%%

\subsection{Scattering amplitude: direct calculation}
\begin{figure}
\includegraphics[width=\columnwidth]{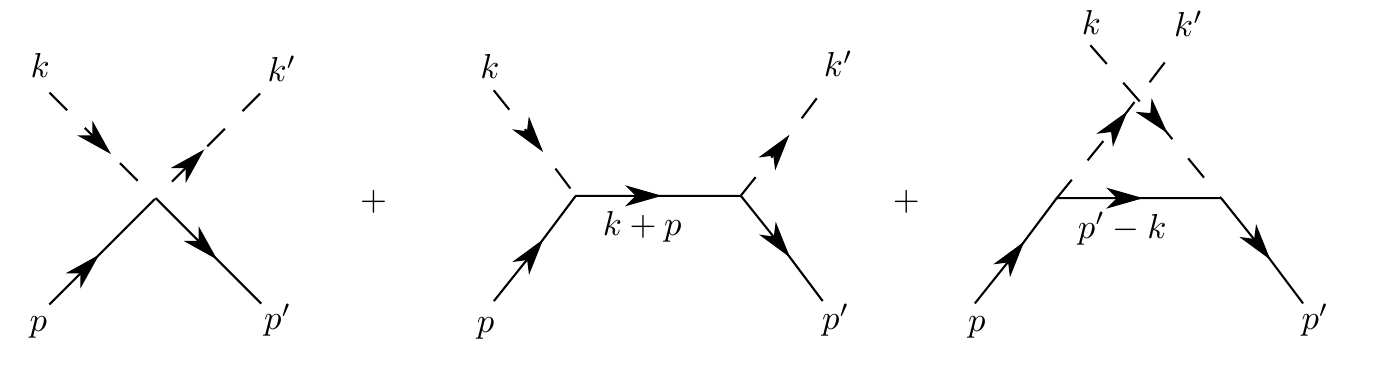}
\caption{Feynman diagrams for the scattering amplitude for the $\pi G\to\pi G$ process. The dashed line stands for the $\pi$ field and the solid line for $G$. The four-momenta of the NG boson in the initial and final state are denoted respectively as $k$ and $k'$, whereas those of the mNG boson are denoted as $p$ and $p'$. The arrows indicate the flow of momentum.}
\label{fig:SO(3)amplitude}
\end{figure}

In order to get insight in the properties of the scattering amplitudes in our model, let us perform a sample calculation and inspect the four-particle process
\begin{equation}
\pi G\to\pi G,
\end{equation}
see Fig.~\ref{fig:SO(3)amplitude} for the corresponding Feynman diagrams and the explanation of our notation. A simple calculation leads to the tree-level result for the on-shell amplitude with amputated external legs~\footnote{This object is what in relativistic field theory is usually called invariant amplitude, referring to its Lorentz invariance. However, since in our case manifest Lorentz invariance is broken by the presence of the chemical potential, we refrain from using this term.},
\begin{equation}
\Ma=\frac2{v^2}k\cdot k'+\frac{2\mu^2k_0k_0'}{v^2}\left(\frac1{p\cdot k}-\frac1{p\cdot k'}\right).
\label{MSO(3)}
\end{equation}
Let us first inspect the properties of this amplitude as the momentum of one of the NG states, say the incoming one, goes to zero. Naively the amplitude vanishes thanks to the presence of the factors of $k$ in the numerators. However, since $p\cdot k'= p'\cdot k$, both terms in the parentheses in Eq.~\eqref{MSO(3)} are singular in this limit. A more careful evaluation leads to
\begin{equation}
\lim_{\vek k\to\vek 0}\Ma=\frac{2\mu^2k_0'}{v^2}\left(\frac1{p_0-|\vek p|\cos\alpha}-\frac1{p_0'-|\vek p'|\cos\beta}\right),
\end{equation}
where $\alpha$ and $\beta$ are the angles between $\vek k$ and $\vek p$ and $\vek p'$, respectively. The absence of Adler's zero in such a scattering process is a well-known issue, which arises from the presence of cubic interaction vertices in the model~\footnote{In fact, it has been shown that in effective theories with derivative couplings such as ours, cubic interaction vertices can always be removed by a field redefinition as long as Lorentz invariance is maintained~\cite{Cheung:2016drk}. As a consequence, the kinematic singularity described in the main text cannot appear in interactions including NG bosons only; it typically arises when a soft NG boson is radiated from a massive, non-NG external particle in the scattering process~\cite{Weinberg:1996v2}. Our model demonstrates that once Lorentz invariance is given up, cubic interaction vertices leading to the kinematic singularity may reappear.}, and thus from the latter two Feynman diagrams in Fig.~\ref{fig:SO(3)amplitude}: as the momentum $k$ goes to zero, the internal propagator in these diagrams approaches the mass shell, leading to a kinematic singularity.

If, on the other hand, one of the mNG bosons in the process becomes soft~\footnote{Such that either $p\xrightarrow{}{(\mu,0)}$ or $p'\xrightarrow{}{(\mu,0)}$}, no such a kinematic singularity appears due to the non-vanishing mass of the mNG boson. A simple manipulation using the kinematics of the process shows that
\begin{equation}
\lim_{\vek p\to\vek0}\Ma=\lim_{\vek p'\to\vek0}\Ma=0.
\end{equation}
This is our first piece of evidence that the interactions of mNG bosons are weak at low momentum in spite of their nonzero mass.

%%%%%%%%%%%%%%%%%%%%%%%%%%%%%%%%%%%%%%%%%%%%%%%%%%%

\subsection{Scattering amplitude from current conservation}

\begin{figure}
\includegraphics[width=\columnwidth]{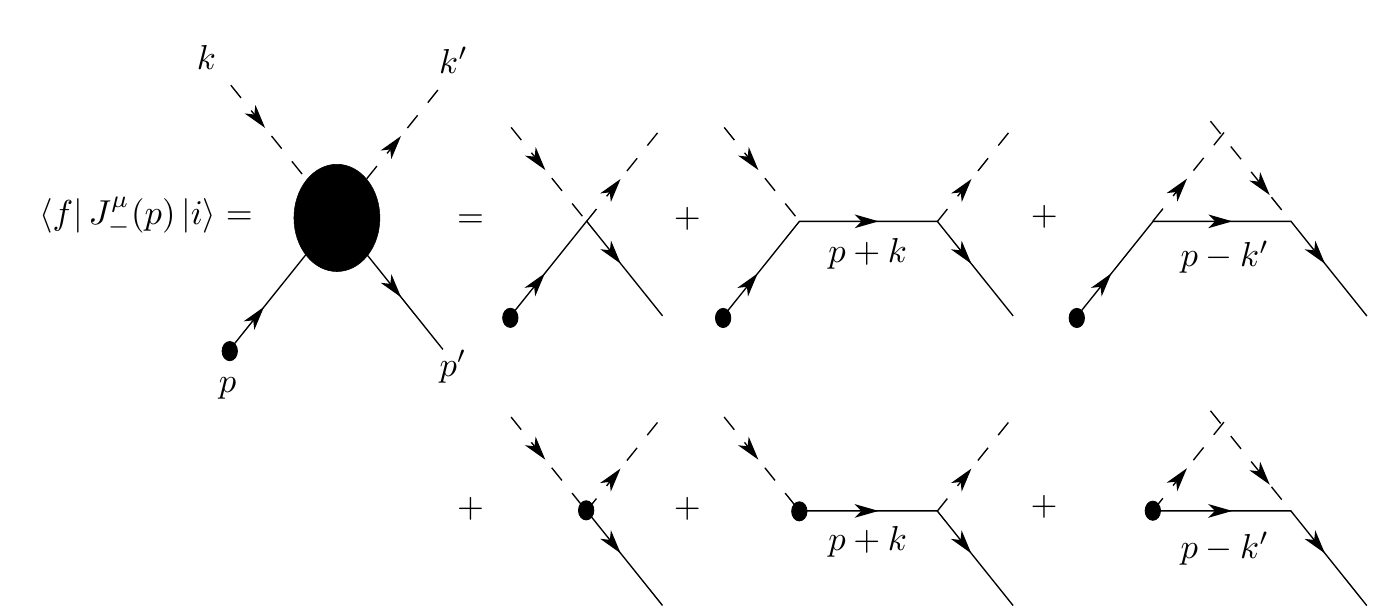}
\caption{Feynman diagrams contributing to the matrix element $\bra fJ_-^\mu(p)\ket i$. The dot on the external line carrying momentum $p$ represents the current operator, otherwise the notation is the same as in Fig.~\ref{fig:SO(3)amplitude}.}
\label{fig:SO(3)current}
\end{figure}

So far, we have found Adler's zero in a single scattering amplitude of the mNG state in our model by a direct computation. Should we be able to prove the existence of Adler's zero for mNG bosons on general grounds, we need a more robust approach. To that end, recall that the presence of Adler's zero for true, massless NG bosons is usually proved as a direct consequence of conservation of the Noether current associated with the spontaneously broken symmetry~\cite{Weinberg:1996v2}. We shall now therefore imagine that the incoming mNG state in the process shown in Fig.~\ref{fig:SO(3)amplitude} is created by the current operator $J^\mu_-$, and investigate the matrix element
\begin{equation}
\bra fJ_-^\mu(p)\ket i\equiv\bra{k',p'}J_-^\mu(p)\ket k.
\label{Jelement}
\end{equation}
Note that the kinematics corresponding to this matrix element is different than that of the scattering amplitude in Fig.~\ref{fig:SO(3)amplitude}: whereas the four-momenta $k,k',p'$ label one-particle asymptotic states and therefore are on-shell, the four-momentum $p$ is created by the local current operator and thus can be off-shell. Keeping this momentum off, if close, the mass shell is of course all-important for understanding the analytical structure of the matrix element and extracting from it the physical scattering amplitude.

As the first step, we write down the perturbative expansion of the Noether currents~\eqref{noetherSO(3)}, just as we previously did for the Lagrangian,
\begin{widetext}
\begin{equation}
\begin{split}
J_1^\mu &= \pi\partial^\mu G - G \partial^\mu \pi + \delta^{\mu 0}\mu G \left(\frac{\pi^2 + G^2}{2v}-v\right) + \dotsb,\\
J_2^\mu &=-\delta^{\mu 0}\mu \pi G -v\partial^\mu G- \frac{1}{2v}\partial^\mu G(G^2-\pi^2)-\frac{1}{v}\pi G \partial^\mu \pi + \dotsb,\\
J_3^\mu &= \delta^{\mu 0}\mu(v^2-G^2) + v\partial^\mu \pi + \frac{1}{v}\pi G \partial^\mu G - \frac{1}{2v}\partial^\mu \pi (G^2-\pi^2) + \dotsb,
\end{split}
\end{equation}
\end{widetext}
where terms of higher order in the fields are omitted.

The matrix element $\bra fJ_-^\mu(p)\ket i$ with $J^\mu_-\equiv J^\mu_1-\imag J^\mu_2$ can now be evaluated perturbatively similarly to the previous direct calculation of the scattering amplitude. The Feynman diagrams that contribute to it are shown in Fig.~\ref{fig:SO(3)current} and fall into two distinct classes. The first three diagrams arise from the part of the current linear in $G$, and contain a pole at $p^2=\mu^2$. The last three arise from the parts of the current quadratic and cubic in the fields, and do not have a simple pole in the $p^2$ variable.

It is obvious from Figs.~\ref{fig:SO(3)amplitude} and~\ref{fig:SO(3)current} that there is a one-to-one correspondence between diagrams contributing to the scattering amplitude for the process $\pi G\to\pi G$ and the pole contributions to the matrix element $\bra fJ_-^\mu(p)\ket i$. Using only the knowledge of the propagator of the $G$-field and of the linear pieces of the current $J^\mu_-$, that is \emph{without} having to evaluate the scattering amplitude explicitly, the pole part of the matrix element of the current can be expressed as
\begin{equation}
\bra fJ_-^\mu(p)\ket i_\text{pole}=-\frac{\imag v(\mu\delta^{\mu0}+p^\mu)}{p^2-\mu^2}(-\imag\Ma_\text{off-shell}),
\end{equation}
where the subscript ``off-shell'' refers to the fact that only the momenta $k,k',p'$ but not $p$ are now on-shell in the scattering amplitude.

The non-pole part of the current matrix element is likewise evaluated straightforwardly,
\begin{equation}
\begin{split}
\bra fJ_-^\mu(p)\ket i_\text{non-pole}={}&\frac1v(\mu\delta^{\mu0}+k^\mu+p'^\mu-k'^\mu)\\
&-\frac{\mu k_0'}{vp'\cdot k'}(\mu\delta^{\mu0}+p^\mu+2k^\mu)\\
&-\frac{\mu k_0}{vp'\cdot k}(\mu\delta^{\mu0}+p^\mu-2k'^\mu),
\end{split}
\end{equation}
where we used momentum conservation and the on-shell condition for $k,k',p'$ to simplify the result.

We shall now see that the scattering amplitude for the $\pi G\to\pi G$ process is actually completely determined by the non-pole diagrams in Fig.~\ref{fig:SO(3)current}. Indeed, the operator momentum conservation condition~\eqref{conservation} implies that
\begin{equation}
(p_\mu-\mu\delta_{\mu0})\bra fJ_-^\mu(p)\ket i=0.
\label{conservationMatEl}
\end{equation}
This leads to a cancellation of the pole in the pole part of the current matrix element, upon which the off-shell amplitude $\Ma_\text{off-shell}$ can be expressed as
\begin{equation}
\Ma_\text{off-shell}=\frac2{v^2}k\cdot k'+\frac{2\mu^2k_0k_0'}{v^2}\left(\frac1{p'\cdot k'}-\frac1{p'\cdot k}\right).
\end{equation}
Once the momentum $p$ is set on the mass shell, this is seen to be equivalent to the previously found result~\eqref{MSO(3)}.

The moral of this exercise is that we do not need to calculate the scattering amplitude explicitly: it can be extracted from the non-pole contributions to the matrix element of the broken current upon using current conservation. This is a major step towards proving that the scattering amplitude vanishes in the limit of zero momentum of the mNG boson. Before proceeding to the general argument, we will however work out in detail another example. In the calculation above, we have namely used heavily the relativistic kinematics to simplify the expressions. We want to see to what extent the situation complicates in systems where not only the interactions, but also the kinematics are not Lorentz-invariant.

%%%%%%%%%%%%%%%%%%%%%%%%%%%%%%%%%%%%%%%%%%%%%%%%%%%

\section{Case study: relativistic model for kaon condensation}
\label{sec:kaon}

Following Ref.~\cite{Miransky:2001tw,*Schafer:2001bq}, we introduce the linear sigma model, defined by the Lagrangian
\begin{equation}
\La=D_\mu\he\phi D^\mu\phi-m^2\he\phi\phi-\lambda(\he\phi\phi)^2,
\end{equation}
where $\phi$ is a doublet of complex scalars and the covariant derivative incorporates a chemical potential via $D_0\phi\equiv(\de_0-\imag\mu)\phi$. The Lagrangian has a manifest $\tilde G=\gr{SU(2)}\times\gr{U(1)}$ symmetry, corresponding to unitary rotations of the $\phi$ doublet. The chemical potential $\mu$ is then associated with the $\gr{U(1)}$ factor of the symmetry group. This model has been used to describe kaon condensation in dense quark matter, where the $\gr{SU(2)}$ stands for isospin and $\gr{U(1)}$ for strangeness.

The full symmetry group of the model in the limit $\mu=0$ is $G=\gr{SO(4)}\simeq\gr{SU(2)}\times\gr{SU(2)}$, which is most easily seen by thinking of $\phi$ as a collection of four real scalar fields. The non-Abelian nature of this symmetry creates a convenient setting for the presence of mNG bosons in the spectrum. When $\mu>m$, the classical ground state of the model carries a nonzero expectation value of $\phi$, and can be chosen as
\begin{equation}
\langle\phi\rangle=\frac1{\sqrt2}\begin{pmatrix}
0\\v
\end{pmatrix},\qquad
v\equiv\sqrt{\frac{\mu^2-m^2}\lambda}.
\end{equation}
The symmetry-breaking pattern then reads
\begin{equation}
\begin{split}
G=\gr{SU(2)}_L\times\gr{SU(2)}_R&\to\gr{SU(2)}',\\
\tilde G=\gr{SU(2)}_L\times\gr{U(1)}_R&\to\gr{U(1)}'.
\end{split}
\end{equation}
Here the primes refer to the fact that the generators of the unbroken $\gr{SU(2)}$ and $\gr{U(1)}$ subgroups are linear combinations of generators of the $\gr{SU(2)}$ and $\gr{U(1)}$ factors in $G$ and $\tilde G$, respectively. We can see that two of the symmetry generators are broken spontaneously and at the same time explicitly by the chemical potential, and thus expect a single mNG boson in the spectrum.

To check this, we parameterize the doublet $\phi$ as
\begin{equation}
\phi\equiv\frac{1}{\sqrt2}\begin{pmatrix}
\vp\\
v+\psi_3+\imag\psi_4
\end{pmatrix},
\label{param}
\end{equation}
where $\vp$ is a complex field, whereas $\psi_{3,4}$ are real. Inserting this into the model Lagrangian and dropping constant terms, it becomes
\begin{align}
\label{LagSU(2)}
\La={}&\de_\mu\vp^*\de^\mu\vp+\imag\mu(\vp^*\de_0\vp-\vp\de_0\vp^*)-\lambda v^2\psi_3^2\\
\notag
&+\frac12(\de_\mu\psi_3)^2+\frac12(\de_\mu\psi_4)^2+\mu(\psi_4\de_0\psi_3-\psi_3\de_0\psi_4)\\
\notag
&-\lambda v\psi_3(2\vp^*\vp+\psi_3^2+\psi_4^2)-\frac\lambda4(2\vp^*\vp+\psi_3^2+\psi_4^2)^2.
\end{align}
It is easy to see that the $\vp$ field excites a pair of states with the dispersion relations
\begin{equation}
\omega_\pm(\vek p)=\sqrt{\vek p^2+\mu^2}\pm\mu.
\label{mNGdisprel}
\end{equation}
These can be thought of as a genuine particle--antiparticle pair thanks to the fact that they carry the charge of the unbroken exact $\gr{U(1)}'$ symmetry. The lighter of the two is gapless and represents a so-called type-B NG boson~\cite{Watanabe:2011ec,*Watanabe:2012hr,*Hidaka:2012ym}. The heavier of the two, on the other hand, has gap $2\mu$. This is the mNG boson of the extended symmetry group $G$, broken both explicitly and spontaneously~\cite{Watanabe:2013uya}. It has been shown by an explicit calculation that its gap does not receive radiative corrections at one loop~\cite{Brauner:2006xm}.

The $\psi_{3,4}$ sector of the model likewise contains two excitations with the nonrelativistic dispersion relations
\begin{equation}
\omega_{3,4}(\vek p)=\sqrt{\vek p^2+3\mu^2-m^2\pm\sqrt{(3\mu^2-m^2)^2+4\mu^2\vek p^2}}.
\end{equation}
One of these modes is gapless and corresponds to a so-called type-A NG boson~\cite{Watanabe:2011ec,*Watanabe:2012hr,*Hidaka:2012ym}. The other one is gapped and represents a Higgs-like mode. In the calculation below, we actually do not need these dispersion relations, but only the propagator in the $\psi_{3,4}$ sector, which takes a matrix form and can be extracted from the bilinear part of the Lagrangian~\eqref{LagSU(2)},
\begin{equation}
\Da(p)=\frac\imag{p^2(p^2-2\lambda v^2)-4\mu^2p_0^2}\begin{pmatrix}
p^2 & -2\imag\mu p_0\\
+2\imag\mu p_0 & p^2-2\lambda v^2
\end{pmatrix}.
\end{equation}
All the other Feynman rules of the model can be read off the Lagrangian~\eqref{LagSU(2)} trivially.

%%%%%%%%%%%%%%%%%%%%%%%%%%%%%%%%%%%%%%%%%%%%%%%%%%%

\subsection{Scattering amplitude: direct calculation}

\begin{figure}
\includegraphics[width=\columnwidth]{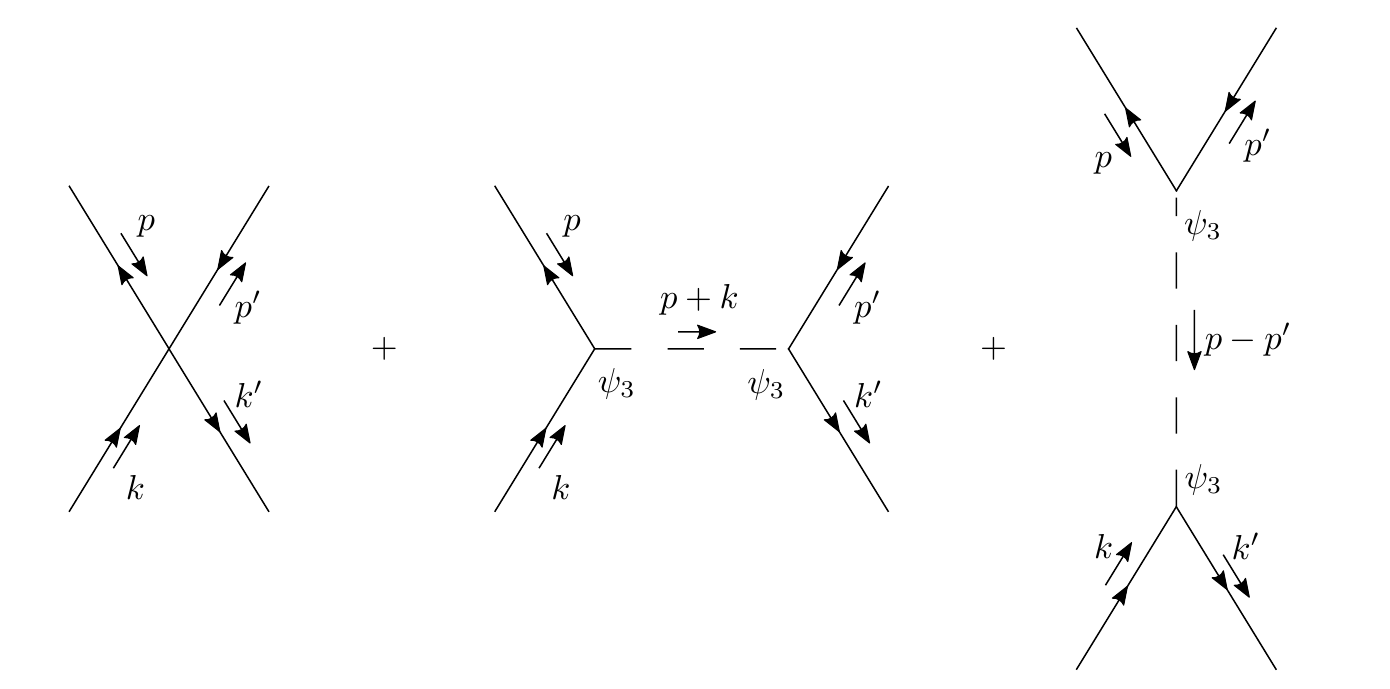}
\caption{Feynman diagrams for the scattering amplitude for the $\text{NG}+\text{mNG}\to\text{NG}+\text{mNG}$ process. All quasiparticles participating in the process are excited by the $\vp$ field. The NG mode is treated as a particle and thus corresponds to an incoming line in the initial state and an outgoing line in the final state. The mNG mode is treated as an antiparticle and thus corresponds to an outgoing line in the initial state and an incoming line in the final state. The dashed line represents the matrix propagator $\Da$; only the $\Da_{\psi_3\psi_3}$ component is needed here since there are no cubic interaction vertices linear in $\psi_4$ in the model. The notation for the four-momenta of the gapless and the gapped state is the same as in Fig.~\ref{fig:SO(3)amplitude}. The arrows on the field lines indicate the flow of the $\gr{U(1)}'$ charge. The flow of momentum is indicated by the arrows next to the momentum labels.}
\label{fig:SU(2)amplitude}
\end{figure}

Let us now, as in the previous section, evaluate the scattering amplitude for a sample scattering process. For the sake of convenience, we choose the process
\begin{equation}
\text{NG}+\text{mNG}\to\text{NG}+\text{mNG},
\end{equation}
where ``NG'' refers to the type-B NG mode of the model, which is the antiparticle of the mNG mode~\footnote{By choosing a process that only involves asymptotic states coupling to the $\vp$ field, we avoid having to deal with mixing in the initial and final state.}. The diagrams contributing to this process at tree level are shown in Fig.~\ref{fig:SU(2)amplitude}, which also explains all the notation required.

A straightforward application of Feynman rules leads to the following intermediate result for the on-shell amplitude with amputated external legs,
\begin{align}
\label{SU(2)result}
\Ma={}&4\lambda\\
\notag
&+\frac{4\lambda^2v^2(p+k)^2}{(p+k)^2[(p+k)^2-2\lambda v^2]-4\mu^2(p_0+k_0)^2}\\
\notag
&+\frac{4\lambda^2v^2(p-p')^2}{(p-p')^2[(p-p')^2-2\lambda v^2]-4\mu^2(p_0-p_0')^2}.
\end{align}
Adler's zero is not manifest in this case, which is common for linear sigma models: a cancellation between two or more Feynman diagrams is usually required in order to ascertain the vanishing of the scattering amplitude in the soft limit. To that end, note that the dispersion relations~\eqref{mNGdisprel} for the NG and mNG mode can be encoded in the kinematic relations
\begin{equation}
p^2=2\mu p_0,\qquad
k^2=-2\mu k_0,
\end{equation}
and analogously for $p'$ and $k'$. It is then easy to see that
\begin{equation}
\begin{split}
(p+k)^2&\xrightarrow{\vek p\to\vek0}2\mu(p_0+k_0),\\
(p-p')^2&\xrightarrow{\vek p\to\vek0}2\mu(p_0-p_0'),
\end{split}
\end{equation}
which immediately leads to the expected result
\begin{equation}
\lim_{\vek p\to\vek0}\Ma=0.
\end{equation}
It is easy to check that in this case, the Adler zero property also holds for the gapless NG mode; there is no kinematic singularity present in this model. That is because of the structure of the cubic interaction vertices: the internal propagator in the diagrams in Fig.~\ref{fig:SU(2)amplitude} carries a different mode than the external legs, and thus remains off-shell in the limit $\vek k\to\vek0$.

%%%%%%%%%%%%%%%%%%%%%%%%%%%%%%%%%%%%%%%%%%%%%%%%%%%

\subsection{Scattering amplitude from current conservation}

\begin{figure}
\includegraphics[width=\columnwidth]{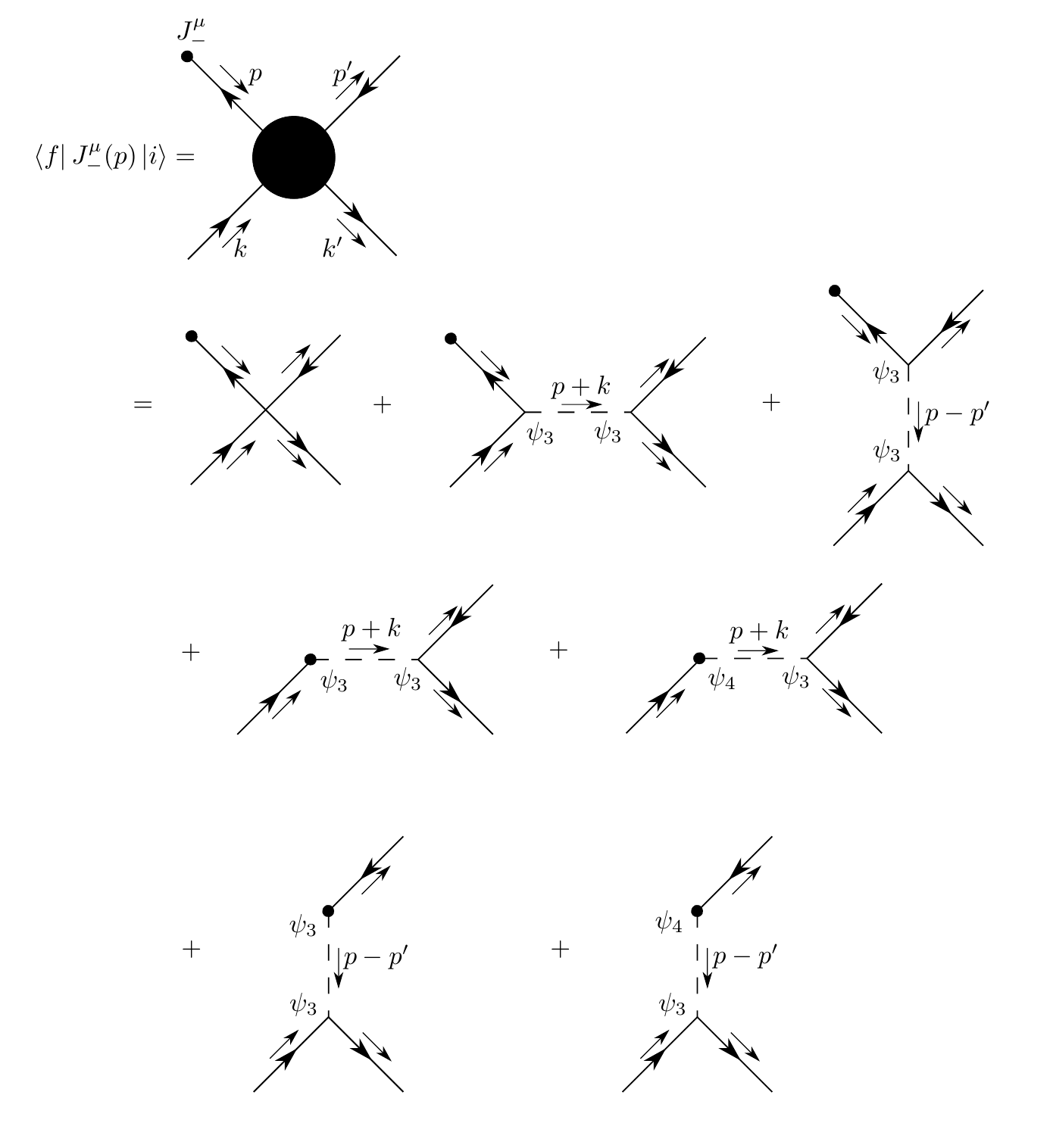}
\caption{Feynman diagrams contributing to the matrix element $\bra fJ_-^\mu(p)\ket i$. The dot on the external line carrying momentum $p$ represents the current operator, otherwise the notation is the same as in Fig.~\ref{fig:SU(2)amplitude}. The $\psi_{3,4}$ labels on the internal propagators indicate that mixing has to be taken into account.}
\label{fig:SU(2)current}
\end{figure}

As the next step, we shall now again see how to reproduce this result without evaluating the scattering amplitude explicitly, using only current conservation. To that end, we first need to identify the Noether current that excites the mNG boson of the model. Adding the chemical potential to the theory explicitly breaks two of the generators of the right $\gr{SU(2)}$ factor in the symmetry group $G$. The corresponding currents take the form
\begin{equation}
\begin{split}
J^\mu_{\text{R}1}&=-\phi^T\tau_2\de_\mu\phi+\he\phi\tau_2\de_\mu\phi^*,\\
J^\mu_{\text{R}2}&=-\imag\phi^T\tau_2\de_\mu\phi-\imag\he\phi\tau_2\de_\mu\phi^*,
\end{split}
\end{equation}
where $\tau_2$ is the second Pauli matrix. In this case, it is more convenient to define the ``ladder currents'' with an additional factor of $\sqrt2$,
\begin{equation}
J^\mu_\pm\equiv\frac1{\sqrt2}(J^\mu_{\text{R}1}\pm\imag J^\mu_{\text{R}2}).
\end{equation}
Only the current $J^\mu_-$ is needed as it excites the mNG boson. Using the parameterization~\eqref{param}, it becomes
\begin{equation}
J^\mu_-=-\imag v\de^\mu\vp-\imag(\psi_3\de^\mu\vp-\vp\de^\mu\psi_3)+(\psi_4\de^\mu\vp-\vp\de^\mu\psi_4).
\end{equation}

As in the previous section, we now want to evaluate the matrix element~\eqref{Jelement}. The Feynman diagrams that contribute to it are displayed in Fig.~\ref{fig:SU(2)current}. The pole part of the matrix element is again related to the scattering amplitude of interest by a simple expression,
\begin{equation}
\bra fJ_-^\mu(p)\ket i_\text{pole}=\frac{\imag vp^\mu}{p^2-2\mu p_0}(-\imag\Ma_\text{off-shell}),
\end{equation}
where the subscript off-shell indicates that only the four-momenta $k$, $k'$ and $p'$ are on-shell. The non-pole part of the matrix element, $\bra fJ_-^\mu(p)\ket i_\text{non-pole}$, which we will for brevity call simply $N^\mu_-$, is now given by a larger number of diagrams as a result of the mixing of the $\psi_{3,4}$ fields. Evaluating all the contributions explicitly yields
\begin{equation}
\begin{split}
N^\mu_-={}&-\frac{2\lambda v(p^\mu+2k^\mu)[(p+k)^2+2\mu(p_0+k_0)]}{(p+k)^2[(p+k)^2-2\lambda v^2]-4\mu^2(p_0+k_0)^2}\\
&-\frac{2\lambda v(p^\mu-2p'^\mu)[(p-p')^2+2\mu(p_0-p_0')]}{(p-p')^2[(p-p')^2-2\lambda v^2]-4\mu^2(p_0-p_0')^2}.
\end{split}
\end{equation}

In the present case where the gap of the mNG mode is $2\mu$, the current conservation condition~\eqref{conservation} implies
\begin{equation}
(p_\mu-2\mu\delta_{\mu0})\bra fJ_-^\mu(p)\ket i=0,
\end{equation}
as opposed to Eq.~\eqref{conservationMatEl}. The prefactor $(p_\mu-2\mu\delta_{\mu0})$ clearly cancels the pole in $\bra fJ_-^\mu(p)\ket i_\text{pole}$, although the propagator of the mNG boson now takes a nonrelativistic form. Upon canceling the pole, the off-shell scattering amplitude can be expressed in terms of the non-pole contributions $N^\mu_-$ as
\begin{equation}
\Ma_\text{off-shell}=-\frac1v(p_\mu N^\mu_--2\mu N^0_-).
\label{Moffshell}
\end{equation}

Upon using some kinematics for the initial and the final state of the scattering process, it is straightforward to show that in the on-shell limit, this result coincides with the previously derived Eq.~\eqref{SU(2)result}. Even more importantly, however, Eq.~\eqref{Moffshell} makes the presence of Adler's zero in the limit $\vek p\to\vek0$ \emph{manifest} as long as $N^\mu_-$ is not singular in this limit, which it is not by construction. [It does not include the contribution of the one-particle pole at $p_0=\omega_+(\vek p)$.] This is the last crucial ingredient that we need for a general proof of the existence of Adler's zero in scattering amplitudes of mNG bosons.

%%%%%%%%%%%%%%%%%%%%%%%%%%%%%%%%%%%%%%%%%%%%%%%%%%%

\section{General argument}
\label{sec:general}

\begin{figure}
\includegraphics[width=\columnwidth]{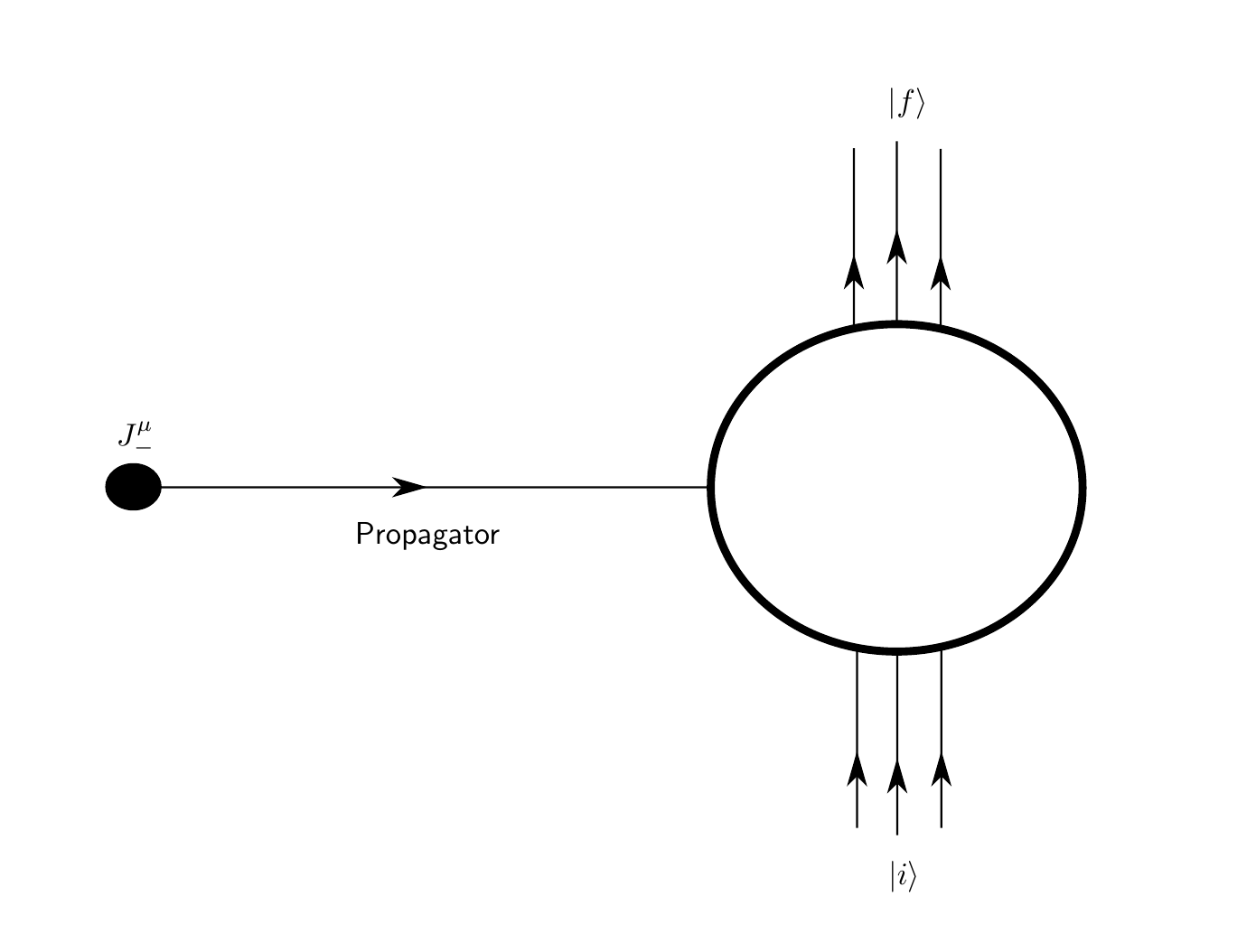}
\caption{A generic scattering process involving a mNG boson. The initial and final state $\ket i$ and $\ket f$ can include an arbitrary number of mNG and non-mNG modes.}
\label{fig:general}
\end{figure}

We would now like to generalize our argument from the previous section so that it:
\begin{itemize}
\item Applies to any (compact) symmetry group and symmetry breaking pattern $G\to H$.
\item Does not require the evaluation of specific Feynman diagrams, but only relies on current conservation.
\item Does not assume any particular form of the propagator of the mNG field.
\end{itemize}
We will follow rather closely the usual proof of existence of Adler's zero for exact spontaneously broken symmetries~\cite{Weinberg:1996v2}. A generic scattering process involving a mNG boson in the initial state can be represented by the diagram in Fig.~\ref{fig:general}. Just like in our above analysis of specific examples, the mNG state is created by a local Noether current operator, and the diagram therefore corresponds to the matrix element $\bra fJ^\mu_-(p)\ket i$, where $p$ is the mNG boson four-momentum.

To understand the analytic structure of this matrix element, we will need the K\"all\'en--Lehmann spectral representation. Its general nonrelativistic version for a time-ordered Green's function of two local fields, $A(x)$ and $B(x)$, takes the form~\cite{Brauner:2006xm}
\begin{align}
\notag
\Da_{AB}(p)={}&\imag(2\pi)^3\sum_n\Biggl[\frac{\bra0A(0)\ket{n,\vek p}\bra{n,\vek p}B(0)\ket0}{p_0-\omega(\vek p)+\imag\eps}\\
\label{kallen}
&-\frac{\bra0B(0)\ket{n,-\vek p}\bra{n,-\vek p}A(0)\ket0}{p_0+\omega(\vek p)-\imag\eps}\Biggr],
\end{align}
where the Hamiltonian eigenstates $\ket{n,\vek p}$ are assumed to be normalized according to $\bra{m,\vek p}n,\vek q\rangle=\delta_{mn}\delta^3(\vek p-\vek q)$ and $\omega_n(\vek p)$ is their energy. Note that the index $n$ is discrete for one-particle states and continuous for multiparticle states. Only the former are relevant for us here.

We now set $A\to\phi$ and $B\to J^\mu_-$, where $\phi$ is an interpolating field for the mNG state, that is a field for which the matrix element $\bra0\phi(0)\ket{G(\vek p)}$ between the many-body vacuum $\ket0$ and the one-particle mNG state $\ket{G(\vek p)}$ is nonzero. The pole part of the two-point function of the current and the interpolating field $\phi$ then reads
\begin{equation}
\Da_{\phi J^\mu_-}(p)\xrightarrow{\text{mNG pole}}\imag(2\pi)^3\frac{\bra0\phi(0)\ket{G(\vek p)}\bra{G(\vek p)}J^\mu_-(0)\ket0}{p_0-\omega(\vek p)},
\end{equation}
where $\omega(\vek p)$ now denotes the dispersion relation of the mNG state. The matrix element $\bra0\phi(0)\ket{G(\vek p)}$ can be naturally absorbed into the definition of the scattering amplitude $\Ma$ of the process, which apart from the initial state $\ket i$ and the final state $\ket f$, also includes a mNG state.

Altogether, the matrix element for the process depicted in Fig.~\ref{fig:general} has the following representation,
\begin{align}
\notag
\bra fJ^\mu_-(p)\ket i={}&\bra{G(\vek p)}J^\mu_-(0)\ket0\frac{\imag(2\pi)^3}{p_0-\omega(\vek p)}(-\imag\Ma_\text{off-shell})\\
\label{finalargument}
&+N^\mu_-(p),
\end{align}
where $N^\mu_-(p)$ is the non-pole contribution. As the next step, we use the parameterization of the current matrix element $\bra{G(\vek p)}J^\mu_-(0)\ket0$, following from Eq.~\eqref{FGdef}, and the current conservation condition~\eqref{conservation}. Some caution is required here: while the four-momentum in Eq.~\eqref{FGdef} is on-shell, that is, the frequency therein equals $\omega(\vek p)$, the four-momentum in Eq.~\eqref{finalargument} is off-shell and its temporal component is denoted simply as $p_0$. Distinguishing carefully the two four-momenta, it is straightforward to see that current conservation leads to a complete cancellation of the pole in Eq.~\eqref{finalargument}, upon which the off-shell scattering amplitude can be expressed as
\begin{equation}
\Ma_\text{off-shell}(p)=\frac{\imag(p_\mu-\mu q\delta_{\mu0})N^\mu_-(p)}{(2\pi)^3\bigl[\omega(\vek p)F_1(|\vek p|)+F_2(|\vek p|)\bigr]}.
\label{finaleq}
\end{equation}
As the final step, we can bring the four-momentum $p$ on-shell and take the soft limit. It is now obvious that the scattering amplitude for the process involving a mNG boson vanishes in the limit $\vek p\to\vek0$ provided that $N^\mu_-$ is not singular in this limit (which it is not by construction), and that the denominator in Eq.~\eqref{finaleq} does not vanish in this limit. That latter requirement is equivalent to the statement that the coupling of the broken charge to the mNG state does not vanish in the soft limit, which is actually one of the hallmarks of mNG bosons~\cite{Watanabe:2013uya}. This concludes our general proof of the existence of Adler's zero in scattering amplitudes of mNG bosons.

%%%%%%%%%%%%%%%%%%%%%%%%%%%%%%%%%%%%%%%%%%%%%%%%%%%

\section{Conclusions}
\label{sec:conclusions}

In this paper, we have analyzed the low-energy properties of scattering amplitudes for processes involving one or more mNG bosons. We showed that as a consequence of exact symmetry constraints, these scattering amplitudes exhibit Adler's zero just like those of ordinary (gapless) NG bosons. When the momentum of the mNG boson is tuned to zero (and the momenta of the other participating particles are modified accordingly to maintain energy and momentum conservation, but otherwise tend to nonzero limits), the scattering amplitude vanishes. There are no kinematic singularities associated with radiation of soft mNG bosons from the initial or final state due to the nonzero gap of the mNG boson.

This result, in fact, ensures that mNG bosons are well-defined quasiparticles in spite of their nonzero gap: due to their weak interactions at low momentum, their width necessarily goes to zero in the long-wavelength limit.

The examples analyzed explicitly in this paper include antiferromagnetic spin waves in an external magnetic field, and a model for kaon condensation in dense quark matter, where the mNG mode is one of the gapped kaons. However, our conclusions hold equally well for other known examples of mNG bosons such as ferromagnetic spin waves in an external magnetic field, or the neutral pion  in the pion superfluid phase of quantum chromodynamics.

%%%%%%%%%%%%%%%%%%%%%%%%%%%%%%%%%%%%%%%%%%%%%%%%%%%

\subsection{Double soft limits of scattering amplitudes}
Given the fact that mNG bosons respect the Adler zero property, it is interesting to consider what happens in the limit where the momenta of \emph{two} NG or mNG bosons tend to zero simultaneously~\footnote{We thank the anonymous referee for asking a question that stimulated the present discussion.}. The behavior of scattering amplitudes of true, gapless NG bosons in this limit has recently attracted considerable attention, see, for instance, Refs.~\cite{ArkaniHamed:2008gz,Low:2015ogb}. The limit of the scattering amplitude in general turns out to be nonzero, and it reflects the non-Abelian nature of the underlying symmetry.

\begin{figure}
\includegraphics[width=\columnwidth]{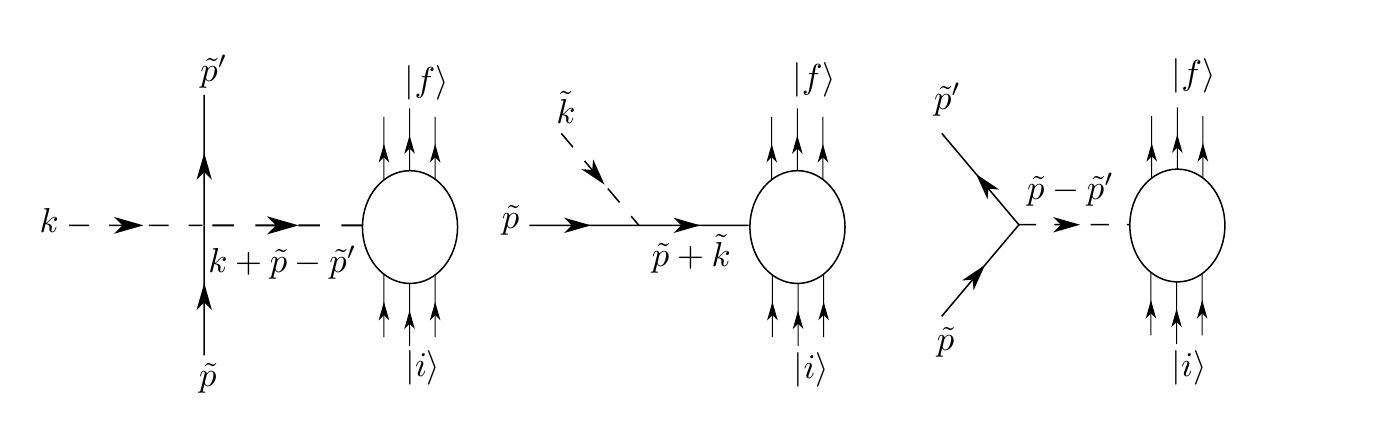}
\caption{Topologies of Feynman diagrams that can potentially lead to a nonzero scattering amplitude in the limit where the momenta of two of the participating particles are sent to zero simultaneously. We use the same notation for lines and vertices as in Sec.~\ref{sec:antiferromagnet}.}
\label{fig:Convert_figure}
\end{figure}

As explained in detail in Ref.~\cite{Low:2015ogb}, this effect arises solely from Feynman diagrams where the two NG bosons in question, and another external leg, are attached to the same quartic interaction vertex; see the first diagram in Fig.~\ref{fig:Convert_figure}. The reason is that when two of the momenta attached to the quartic vertex go to zero, the propagator attached to it becomes on-shell, and the resulting singularity may cancel the suppression of the amplitude due to the presence of derivatives in the vertex.

In order to see a singularity in processes involving two (m)NG bosons interacting through such a quartic vertex, it is essential that both momenta and energies of the two modes add up to zero in the soft limit. This excludes a nontrivial double soft limit in processes involving one NG and one mNG boson, and in processes involving two mNG bosons in the initial or final state. The only possibility seems to be processes where one of the mNG bosons is in the initial and the other in the final state.

For illustration, let us recall the effective theory for antiferromagnets, discussed in Sec.~\ref{sec:antiferromagnet}. Following the notation introduced therein, we write the four-momenta of the incoming and outgoing mNG boson including a scaling factor $z$ as
\begin{equation}
\begin{split}
\tilde p^\mu&=(\sqrt{\mu^2+z^2\vek p^2},z\vek p)=(\mu+\tfrac{z^2\vek p^2}{2\mu},z\vek p)+\dotsb,\\
\tilde p'^\mu&=(\sqrt{\mu^2+z^2\vek p'^2},z\vek p')=(\mu+\tfrac{z^2\vek p'^2}{2\mu},z\vek p')+\dotsb,
\end{split}
\end{equation}
where the ellipsis stands for terms of order $z^4$ or smaller. Using the Feynman rules following from the Lagrangian~\eqref{ferroLag}, the first diagram in Fig.~\ref{fig:Convert_figure} evaluates to
\begin{equation}
\frac{\imag}{v^2}(\tilde p-\tilde p')^2\frac{\imag(-\imag\Ma)}{(k+\tilde p-\tilde p')^2}=\frac{\imag z\Ma}{2v^2}\frac{(\vek p-\vek p')^2}{\vek k\cdot(\vek p-\vek p')}+\mathcal O(z^2),
\end{equation}
where $-\imag\Ma$ is the amplitude corresponding to the blob in the diagram. We can see that in this concrete example, the double soft limit of the full scattering amplitude is safe. However, in general we expect diagrams with this topology to give a nontrivial limit when the momenta of one incoming and one outgoing mNG boson go to zero simultaneously.

Next, let us have a look at the second diagram in Fig.~\ref{fig:Convert_figure}. This type of kinematics was already observed in Sec.~\ref{sec:antiferromagnet} to lead to a violation of the Adler zero property for the gapless NG boson. What if now the momentum of the incoming mNG boson goes to zero as well? Setting $\tilde k^\mu=zk^\mu$, a simple calculation gives for this diagram,
\begin{equation}
-\frac{2\mu}v\tilde k_0\frac{\imag(-\imag\Ma)}{(\tilde p+\tilde k)^2-\mu^2}=-\frac\Ma v+\mathcal O(z).
\end{equation}
In this case, we do get a nonzero double soft limit. That is, however, not so surprising given the fact that already the radiation of the soft gapless NG boson from the external mNG boson line makes the scattering amplitude nonzero at low momentum.

To see a truly new effect, only existing in presence of mNG bosons, consider finally the last diagram in Fig.~\ref{fig:Convert_figure}. As in the case of the diagram with a quartic vertex, we assume that one of the mNG bosons shown in the figure is incoming, while the other is outgoing. We then get
\begin{equation}
\frac{2\mu}v(\tilde p_0-\tilde p_0')\frac{\imag(-\imag\Ma)}{(\tilde p-\tilde p')^2}=-\frac\Ma{\mu v}\frac{\vek p^2-\vek p'^2}{(\vek p-\vek p')^2}+\mathcal O(z^2).
\end{equation}
This kind of nonzero double soft limit arising from a cubic interaction vertex cannot appear in Lorentz-invariant theories for gapless NG bosons, as such cubic vertices can be removed from the theory altogether by a field redefinition~\cite{Cheung:2016drk}.

Altogether, we have identified three different mechanisms whereby a nontrivial double soft limit of scattering amplitudes may be realized in theories with mNG bosons. The first one appears when a NG boson and a mNG boson in the initial or final state are attached to the same cubic interaction vertex. This case accompanies the violation of the Adler zero property for the gapless NG boson alone. The second and third mechanism are both associated with a pair of mNG bosons, one in the initial and the other in the final state of the scattering process. Whether they are attached to a cubic or a quartic vertex, their presence leads to a singular propagator in the Feynman diagram and thus potentially a nonvanishing soft limit of the scattering amplitude.

%%%%%%%%%%%%%%%%%%%%%%%%%%%%%%%%%%%%%%%%%%%%%%%%%%%

\subsection{Scattering amplitudes of pseudo-NG bosons}

What we have not touched upon so far was the scattering amplitudes of pNG bosons that are not mNG bosons, yet their mass also arises from the chemical potential in the system. As mentioned in the introduction, this is a somewhat more exotic, yet perfectly viable possibility. One might expect equally strong constraints on the scattering amplitudes in this case, since after all, we still have the exact conservation law~\eqref{conservation}. However, it is known that the properties of such pNG bosons differ from those of the mNG bosons. Apart from the obvious fact that their gap is not determined by the symmetry and chemical potential alone, they also couple differently to the broken current: unlike for the true mNG bosons, this coupling vanishes in the limit of low momentum~\cite{Watanabe:2013uya}, which invalidates our proof of the existence of Adler's zero in Sec.~\ref{sec:general} for the case of pNG bosons. 

Based on this observation, we conjecture that the scattering amplitudes of pNG bosons whose mass arises from the chemical potential do \emph{not} have the Adler zero property, just like the amplitudes of any other pNG bosons. In order to test this conjecture, we have analyzed to some extent a model where a global $\gr{SO(3)}$ symmetry is completely broken. It is known that in presence of a chemical potential for one of the generators, this system has one NG, one mNG, and one pNG boson~\cite{Watanabe:2013uya,Nicolis:2013sga}. To our surprise, the scattering amplitude for the process we chose to analyze still exhibits Adler's zero. However, our general argument given in Sec.~\ref{sec:general} does not apply to this case, and a further, more detailed investigation is therefore required. We leave this issue to the future. For the sake of convenience, we provide some details of our preliminary analysis in the appendix.

%%%%%%%%%%%%%%%%%%%%%%%%%%%%%%%%%%%%%%%%%%%%%%%%%%%

\acknowledgments
The authors would like to express their thanks to Jens Oluf Andersen for numerous discussions of the subject. This work has been supported in part by a grant within the ToppForsk-UiS program of the University of Stavanger and the University Fund.

%%%%%%%%%%%%%%%%%%%%%%%%%%%%%%%%%%%%%%%%%%%%%%%%%%%

\appendix

\section{Example of non-mNG-type pNG boson}
\label{app:O(3)model}

In this appendix, we shall analyze a low-energy effective theory for a complete spontaneous breaking of an $\gr{SO(3)}$ symmetry. To that end, we shall use the effective Lagrangian formalism, developed in Ref.~\cite{Leutwyler:1993gf,*Leutwyler:1993iq,*Watanabe:2014fva,*Andersen:2014ywa}, whose notation we closely follow.

The leading-order effective Lagrangian for NG bosons in a relativistic system in presence of background gauge fields reads
\begin{equation}
\La=\frac12g_{ab}(\pi)D_\mu\pi^a D^\mu\pi^b.
\label{O(3)lag}
\end{equation}
Here $\pi^a$ are the NG fields that parameterize the coset space of broken symmetry, $G/H$. The Latin indices $a,b,\dotsc$ label broken generators from this coset space. In contrast, the Latin indices $i,j,\dotsc$ will denote generic generators of the whole symmetry group $G$. There is one external gauge field $A^i_\mu$ assigned to each generator $T_i$, and it enters the covariant derivative of the NG field via
\begin{equation}
D_\mu\pi^a\equiv\de_\mu\pi^a-A^i_\mu h^a_i(\pi),
\end{equation}
where $h^a_i(\pi)$ are the Killing vectors that realize the action of the symmetry group $G$ on the coset space $G/H$. Finally, the object $g_{ab}(\pi)$ in Eq.~\eqref{O(3)lag} is a $G$-invariant metric on the coset space, which is determined uniquely up to a set of a priori unknown parameters, which represent the low-energy couplings of the effective theory.

The invariant metric can be determined directly in terms of the Maurer--Cartan form $\omega^i_a(\pi)$, defined by
\begin{equation}
\omega^i_a(\pi)T_i\equiv-\imag U(\pi)^{-1}\frac{\de U(\pi)}{\de\pi^a},
\end{equation}
where $U(\pi)$ is a representative element of the coset space $G/H$, which encodes the NG fields $\pi^a$. Imposing the $G$-invariance of the Lagrangian, we obtain
\begin{equation}
g_{ab}(\pi)=g_{cd}(0)\omega^c_a(\pi)\omega^d_b(\pi),
\label{gabpi}
\end{equation}
where $g_{cd}(0)$ is a set of constants that play the role of the low-energy effective couplings; their values are constrained by the requirement that $g_{ab}(0)$ be a symmetric invariant tensor of the unbroken subgroup $H$. Eq.~\eqref{gabpi} makes it clear that we do not really need to know the full Killing vectors $h^a_i(\pi)$, but only their projections of the form $\omega^c_a(\pi)h^a_i(\pi)=\nu^c_i(\pi)$, where the rotation matrix $\nu^i_j(\pi)$ is defined by
\begin{equation}
\nu^i_j(\pi)T_i\equiv U(\pi)^{-1}T_jU(\pi).
\end{equation}
The above relations determine completely the structure of the leading-order effective Lagrangian for an arbitrary symmetry-breaking pattern $G/H$.

%%%%%%%%%%%%%%%%%%%%%%%%%%%%%%%%%%%%%%%%%%%%%%%%%%%

\subsection{Effective Lagrangian and the spectrum}

Let us now see how the above general formalism applies to the case where the continuous $\gr{SO(3)}$ rotation symmetry is completely broken. Without loss of generality, we can assume that the matrix $g_{ab}(0)$ of effective couplings has a diagonal form,
\begin{equation}
g_{ab}(0)\equiv\text{diag}(g_1,g_2,g_3).
\end{equation}
We will turn on a chemical potential for the third generator of $\gr{SO(3)}$, that is, set
\begin{equation}
A^i_\mu=\delta_{\mu0}\delta^{i3}\mu.
\end{equation}
This determines the effective Lagrangian completely via Eq.~\eqref{O(3)lag}. For the moment, we will only need the part of the Lagrangian bilinear in the NG fields $\pi^a$, which is, up to a rescaling of the fields, independent of the choice of parameterization of the matrix $U(\pi)$,
\begin{equation}
\begin{split}
\La_\text{bilin}={}&\frac12\bigl[g_1(\de_\mu\pi_1)^2+g_2(\de_\mu\pi_2)^2+g_3(\de_\mu\pi_3)^2\bigr]\\
&+\frac12\mu(g_1+g_2-g_3)(\pi_1\dot\pi_2-\pi_2\dot\pi_1)\\
&-\frac12\mu^2(g_3-g_2)\pi_1^2-\frac12\mu^2(g_3-g_1)\pi_2^2.
\end{split}
\end{equation}
The form of the mass terms indicates that the ground state is stable under the perturbation caused by the chemical potential provided that $g_3$ is larger than both $g_1$ and $g_2$, which we will from now on assume.

The excitation spectrum of the theory is easy to work out. First, the $\pi_3$ mode does not feel the presence of the chemical potential, and thus behaves as an ordinary gapless NG boson: its dispersion relation reads
\begin{equation}
\omega_3(\vek p)=|\vek p|.
\end{equation}
The $\pi_{1,2}$ modes mix and their dispersion relations therefore take a more complicated form,
\begin{equation}
\omega_\pm(\vek p)^2=\vek p^2+\mu^2+\frac{g_3(g_3-g_1-g_2)}{2g_1g_2}\mu^2(1\pm\Omega_{\vek p}),
\end{equation}
where
\begin{equation}
\Omega_{\vek p}\equiv\sqrt{1+\frac{4g_1g_2}{g_3^2}\frac{\vek p^2}{\mu^2}}.
\end{equation}
From here, we can in turn extract the mass spectrum in the $\pi_{1,2}$ sector,
\begin{equation}
m_+=\mu\sqrt{\frac{(g_3-g_1)(g_3-g_2)}{g_1g_2}},\qquad
m_-=\mu.
\end{equation}
Whereas we find one mNG mode as predicted by Eq.~\eqref{counting}, there is also another pNG mode which is not of the mNG type, although its mass comes from the chemical potential alone. It is this mode that is of interest to us.

%%%%%%%%%%%%%%%%%%%%%%%%%%%%%%%%%%%%%%%%%%%%%%%%%%%

\subsection{Coupling of fields to states}

The analysis of scattering amplitudes in the present model is complicated by the mixing in the $\pi_{1,2}$ sector. In such a situation, it is mandatory to use the Lehmann--Symanzik--Zimmermann formalism to extract the physical scattering amplitude from the off-shell Green's function of the fields. To that end, we need to know how the fields couple to the asymptotic one-particle states in the scattering process.

Such coupling can be extracted from the propagators of the fields using the K\"all\'en--Lehmann spectral representation~\eqref{kallen}. The propagator of $\pi_3$ in the interaction picture is just that of a free massless scalar field, and we readily obtain
\begin{equation}
\bra0\pi_3(0)\ket{3,\vek p}=\frac1{\sqrt{(2\pi)^32g_3|\vek p|}}.
\end{equation}
To extract the couplings between the fields $\pi_{1,2}$ and the states $\ket{\pm,\vek p}$ with the dispersion relations $\omega_\pm(\vek p)$, we first write down the matrix inverse propagator in the $\pi_{1,2}$ sector, following from the Lagrangian $\La_\text{bilin}$,
\begin{equation}
\Da^{-1}(\omega,\vek p)=\begin{pmatrix}
g_1p^2-\mu^2(g_3-g_2) & -\imag\mu\omega(g_1+g_2-g_3)\\
+\imag\mu\omega(g_1+g_2-g_3) & g_2p^2-\mu^2(g_3-g_1)
\end{pmatrix}.
\end{equation}
By looking in turn at the poles at $\omega=\omega_\pm(\vek p)$ and using the spectral representation~\eqref{kallen}, we then find
\begin{equation}
\begin{split}
\bra0\pi_1(0)\ket{+,\vek p}&=\frac1{(2\pi)^{3/2}}\sqrt{\frac{\frac{g_3}{2g_1}(\Omega_{\vek p}+1)-1}{2g_3\Omega_{\vek p}\omega_+(\vek p)}},\\
\bra0\pi_2(0)\ket{+,\vek p}&=\frac\imag{(2\pi)^{3/2}}\sqrt{\frac{\frac{g_3}{2g_2}(\Omega_{\vek p}+1)-1}{2g_3\Omega_{\vek p}\omega_+(\vek p)}},\\
\bra0\pi_1(0)\ket{-,\vek p}&=\frac1{(2\pi)^{3/2}}\sqrt{\frac{\frac{g_3}{2g_1}(\Omega_{\vek p}-1)+1}{2g_3\Omega_{\vek p}\omega_-(\vek p)}},\\
\bra0\pi_2(0)\ket{-,\vek p}&=\frac{-\imag}{(2\pi)^{3/2}}\sqrt{\frac{\frac{g_3}{2g_2}(\Omega_{\vek p}-1)+1}{2g_3\Omega_{\vek p}\omega_-(\vek p)}}.
\end{split}
\end{equation}

%%%%%%%%%%%%%%%%%%%%%%%%%%%%%%%%%%%%%%%%%%%%%%%%%%%

\subsection{Evaluation of scattering amplitudes}

The evaluation of the scattering amplitude for a given process proceeds according to the following steps:
\begin{itemize}
\item Choose a specific parameterization of the matrix $U(\pi)$ and expand the Lagrangian up to the desired order in the fields $\pi^a$.
\item Extract the interaction vertices from the expanded Lagrangian.
\item Construct all tree-level Feynman diagrams contributing to the given process. Note that as a result of the mixing in the $\pi_{1,2}$ sector, diagrams with different fields attached to the external legs may contribute to the same process, since different fields couple to the same one-particle state~\cite{Brauner:2006xm}.
\item Test scaling of the scattering amplitude in the long-wavelength limit numerically.
\end{itemize}
The last point deserves a more detailed comment. Already for four-particle scattering, a relatively large number of Feynman diagrams may contribute as a result of the mixing, which makes testing the asymptotic behavior of the scattering amplitude in the long-wavelength limit analytically difficult. It is more convenient to perform a numerical ``experiment''~\cite{Cheung:2014dqa}. All one needs to do is to generate a set of random kinematical variables that satisfy the energy and momentum conservation conditions for a given process. One then introduces a scaling parameter $z$ into the momentum of the particle whose soft limit is to be investigated, and makes sure that the momenta of all other participating particles are modified so that the on-shell and conservation conditions are satisfied for any value of $z$. Finally, one simply plots the value of the scattering amplitude as a function of $z$ as $z$ tends to zero.

In this way, we have verified that the scattering amplitudes of the mNG boson ($\omega_-$) of the model exhibit Adler's zero as expected, using the $\text{NG}+\text{mNG}\to\text{NG}+\text{mNG}$ process as an example. Then we analyzed analogously the $\text{NG}+\text{pNG}\to\text{NG}+\text{pNG}$ process. Surprisingly, the scattering amplitude still vanishes as the momentum of one of the pNG bosons tends to zero. This might be a special property of the process that we chose to study, or due to some hidden symmetry of the model at hand that we are not aware of.

This issue would definitely deserve a more careful look. While we do not show the details of our evaluation of the scattering amplitudes as they are specific for the chosen parameterization of $U(\pi)$ and the chosen set of random kinematical variables, we do hope that the details presented in this appendix will enable others to reproduce our results, and go beyond.

%%%%%%%%%%%%%%%%%%%%%%%%%%%%%%%%%%%%%%%%%%%%%%%%%%%

\bibliography{references}

%%%%%%%%%%%%%%%%%%%%%%%%%%%%%%%%%%%%%%%%%%%%%%%%%%%

\end{document}